\documentclass{article} 
\usepackage{amsmath}  
\usepackage{multicol}  
\usepackage{graphicx}   
\usepackage{ragged2e}   
\usepackage{amsthm}
\usepackage{booktabs}
\usepackage{array}
\usepackage{float}   
\usepackage{caption}     
\usepackage{subfigure}   
\usepackage{fancyhdr}
\usepackage{indentfirst} 
\usepackage{geometry}
\usepackage{cuted}
\usepackage{setspace}
\usepackage{listings}   
\usepackage{xcolor}
\usepackage{url}
\usepackage{chngcntr}
\counterwithin{figure}{section}
\counterwithin{table}{section}
\begin{document}  
\title{\LARGE\bf Research on restaurant recommendation using machine learning}	
\author{Junan Pan, Zhihao Zhao}
\date{}		
\maketitle     

\begin{spacing}{1.5}
\large{\bf Abstract:}    A recommender system is a system that helps users filter irrelevant information and create user interest models based on their historical records. With the continuous development of Internet information, recommendation systems have received widespread attention in the industry. In this era of ubiquitous data and information, how to obtain and analyze these data has become the research topic of many people. In view of this situation, this paper makes some brief overviews of machine learning-related recommendation systems. By analyzing some technologies and ideas used by machine learning in recommender systems, let more people understand what is Big data and what is machine learning. The most important point is to let everyone understand the profound impact of machine learning on our daily life.
     

 \section{Introduction}
 
 With the development of science and technology, people gradually turn their attention to the field of data mining. Various data mining techniques are utilized in real life. Machine learning technology has played a big role in it. The most widely known is the "Battle of the Century" between AlphaGo and Li Shishi, known as the man-machine war, and finally ended with the victory of AlphaGo, once again showing people machine learning of power. In the eyes of many people, machine learning is a very abstract concept, and this article will analyze it in the field of recommendation systems.
 
 \section{The process of machine learning}
 What is machine learning? In layman's terms, machine learning is to let machines learn and summarize "experience" like humans. Of course, machines cannot accumulate "experience" by going through various things like humans. Instead, let the machine analyze the existing data, summarize the rules, and summarize to form a set of models, which can be applied to real life.
\subsection{Get data} 
 The first step in machine learning is to obtain data. Machine learning without data is empty talk. The acquisition of data is easier than the later steps, because data is everywhere. There are purchase records of consumers in supermarkets, driving records of vehicles on driving recorders, and viewing records of movie fans in movie theaters. These data information is the premise for us to start machine learning.
 \subsection{Clean data}
 The data we obtain is irregular, there is a lot of redundant data that is not useful to us, and there may be misinformation that affects our analysis results. We need to preprocess this information before conducting specific analysis, so as not to affect the accuracy of the analysis results. Data recommendation system based on machine learning is very common in this era of ubiquitous data and information, how to obtain and analyze these data has become the research topic of many people. In view of this situation, this paper makes some brief overviews of machine learning-related recommendation systems. By analyzing some technologies and ideas used by machine learning in recommender systems, let more people understand what is big data and what is machine learning.The most important point is to let everyone understand the profound impact of machine learning on our daily life.

Abstract Preprocessing is mainly divided into three aspects: distance measurement, sampling, dimensionality reduction.

The KNN classification (k-NearestNeighbor) used in collaborative filtering recommender systems mainly depends on the distance metric method. The more commonly used distance measurement methods are Euclidean distance, Pearson correlation coefficient, Jaccard coefficient (for binary attributes), etc. Sampling is the main technique for data mining to select relevant data subsets from large data sets. It also plays an important role in the final interpretation step. The most commonly used sampling method is non-replacement sampling. Taken from the population, but it is also permissible to perform substitution sampling, that is, the item does not need to be removed from the population even if it is selected. Usually the ratio of training and test sets is 80/20. The final dimension reduction is to remove some points that are very sparse and have little impact on the result set, reduce the dimension, avoid dimension disaster, and reduce the difficulty of operation.

\subsection{Analyzing data and building models}
In the whole process of machine learning, the most difficult and core is to analyze data. There are many ways to analyze data, each of which has different effects in different practical applications, and needs to be analyzed on a case-by-case basis. Here are some of the more commonly used data analysis methods in recommender systems:

Nearest Neighbor Algorithm (KNN): KNN predicts the label class of unknown samples by storing training records and using them. Such classifiers store the entire training set and only classify new records when they exactly match the training set. Compared with other machine learning algorithms, KNN is the simplest, because KNN does not need to build an explicit model, which is called a lazy learner. Although the KNN method is simple and intuitive, its results are accurate and very easy to improve.

k-means algorithm: The k-means algorithm is a block clustering algorithm that divides the acquired n object data into k uncorrelated subsets (k$\textless$n). It is similar to the expectation-maximization algorithm that deals with mixed normal distributions in that they both try to find the centers of natural clusters in the data. It assumes that object attributes come from spatial vectors, and the goal is to minimize the sum of mean squared errors within each group. The k-means algorithm starts by randomly selecting k center points, and all items are assigned to the class of their closest center node. When items are newly added or removed, the central node of the new cluster needs to be updated, and the membership of the cluster also needs to be updated, and so on and so forth until no items change their cluster membership. The final cluster is very sensitive to the initial center point, and there may also be empty clusters.

Artificial Neural Network (ANN): The artificial neural network algorithm simulates a biological network, which consists of a set of internal connection points and weighted chains, and is a type of pattern matching algorithm. Often used to solve classification and regression problems. ANN is a huge branch of machine learning, there are hundreds of different algorithms, and deep learning is one of its important components. The main advantage of ANN is that it can handle nonlinear classification tasks, and through parallel processing, it can operate in the case of partial network damage. But it is difficult for ANN to provide an ideal network topology for a given problem.When the topology is determined, its performance level is at the lower line of the classification error rate.

Bayesian Classifiers: Bayesian classifiers are a class of algorithms based on the definition of probability and Bayes' theorem, the school of Bayesian statistics that uses probability to represent uncertainty in relationships learned from data. It treats each attribute and class label as a random variable. Given a record with N attributes ($A_1 , A_2 , A_3 , ..., A_N$ ), the goal is to predict a class $C_k$ , by making a , ..., $A_N$ ), find the value of $C_k$ that maximizes the posterior probability for that class. Common Bayesian classifier algorithms include Naive Bayesian Algorithm, Average Single Dependency Estimation (AODE), and Bayesian Network (BBN).

\subsection{Test the model}
The last step in the whole machine learning is to test the model and check the accuracy of the model. This is an important step in measuring the merits of an algorithm. The test data set can be randomly selected from the test or obtained from the test set reserved in advance.

\section{Construction of restaurant recommendation system based on deep learning}
\subsection{Restaurant matching recommendation system based on deep learning}
\subsubsection{Overall Framework}
The overall framework is shown in Figure 3.1. First, the recall set is filtered out according to the given information. Specifically, the user's historical purchase data is used. Usually, a user buys two clothing products at the same time in a short period of time. The probability of matching will increase; secondly, using the text information of the product title, the recall set of the matching product to be predicted is obtained by calculating the cosine similarity. After the recall set is obtained, it is filtered by collocation categories and weighted for fusion according to their respective hit rates. Finally, the features are obtained through the convolutional neural network, and the logistic regression model is used to calculate the matching probability between the recall set and the test set, and use this probability to revise and reorder the fusion recall set.
\begin{figure}[H]
  \centering
  \includegraphics[width=14cm]{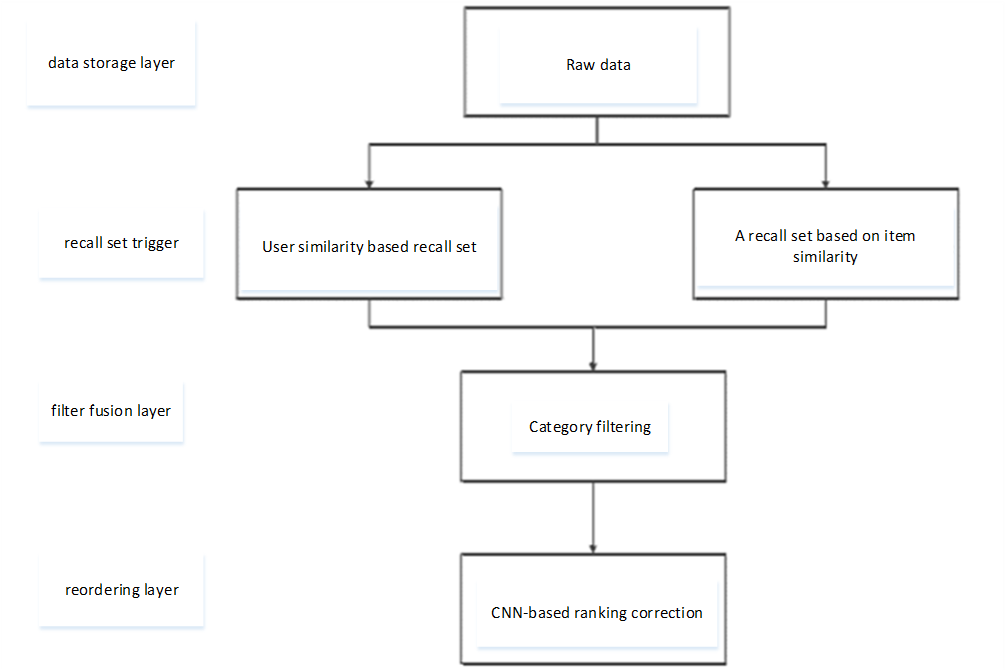}
  \caption{The overall framework of a recommendation system based on deep learning}
\end{figure}

\subsubsection{Implementation of recall set triggering based on MapReduce}
Mapreduce is a distributed computing framework, mainly used to process large-scale data, which was proposed by Google in 2003. Its idea draws on the map function and reduce function of functional programming, and then the open source community Apache implements the Hadoop ecosystem, and its feature of supporting MapReduce makes it rapidly and widely used in engineering practice. After that, Alibaba Group realized the ODPS platform, and its support for MapReduce and elastic cloud computing features also attracted quite a lot of developers.

MapReduce is mainly divided into two stages, namely map and reduce. The map task is mainly responsible for data loading, parsing, transformation and filtering. The reduce task is responsible for processing a subset of the map task output. The entire processing flow is shown in Figure 3.2.
\begin{figure}[H]
  \centering
  \includegraphics[width=16cm]{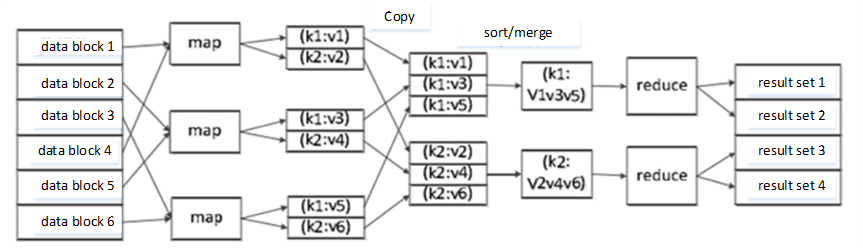}
  \caption{MapReduce processing flow}
\end{figure}

The recall set based on user similarity is mainly to obtain the number of people who have purchased any two products at the same time in a short period of time. The formula is expressed as follows:
 
Among them, (I$\cdot$) is the indicator function, when the function body is true, the count is increased by 1, It is any commodity, Bu is the set of commodities purchased by user u, Tui is the time when user u purchased commodity i, and $\tau$ is the time interval . This formula needs to be solved by two-step MapReduce. The specific solution process is as follows:
$$
{sim}\left(I t_{i}, I t_{j}\right) \approx {freq}\left(I t_{i}, I t_{j}\right)=I_{u \in U}\left(I t_{i} \in B_{u} \cap I t_{j} \in B_{u} \mid T_{u i}-T_{u j}<\tau\right)
$$

(1) Read the original data, use the user id as the key in the map stage, and output the product id and purchase date as the value; (2) in the reduce stage, combine all the products of a user whose purchase time difference does not exceed $\tau$ months, 
and use Commodity ids are sorted and output in descending order; (3) In the second map stage, the pair of commodity id pairs are used as keys, and 1 is used as value for output; 
(4) In the reduce stage, all the same id pairs are added up, and the final output, Product id pair and the number of purchasers. The construction of recall set based on product similarity is a little more complicated. It uses Natural Language Processing (NLP). To get the text similarity of product titles, the $T_f-I_{df}$ weight of each document word segmentation must be obtained first. The process is shown in Figure 3.3. shown. Among them, $T_f-I_{df}$ (Termfrequency–Inversedocumentfrequency) is a commonly used weighting technique for data mining. Its calculation is divided into two steps, namely the weight of term frequency (TermFrequency, TF) and the weight of inverse document frequency (InverseDocumentFrequency, IDF).
 \begin{figure}[H]
  \centering
  \includegraphics[width=16cm]{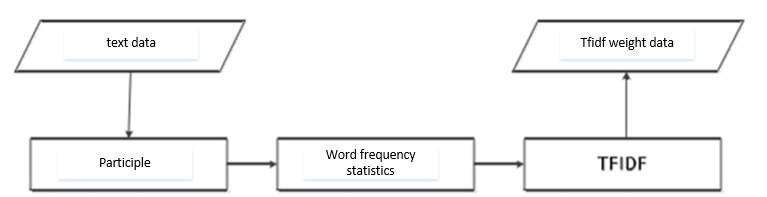}
  \caption{$T_f-I_{df}$ calculation process}
\end{figure}
 
The word frequency is weighted according to the frequency of the word segmentation in the document. The formula is as follows:
 $$
\mathrm{w}_{k j}=t f_{k j}
$$

Among them, $tf_{kj}$ is the word frequency of the participle $t_k$ in the document $d_j$.

The inverse document frequency is to solve a problem in the weight of word frequency, specifically, all word segmentations are considered to be equally important. However, if all documents contain the same word, then the word segmentation has no degree of discrimination, and the original document cannot be distinguished from others. Then the inverse document frequency can be expressed as:
$$
\mathrm{i} d f_{k}=\log \frac{m}{n_{k}}
$$

Among them, m is the number of all documents, and $n_k$ is the number of documents in the document set D of the segmented word tk. According to Equation 4-5, if a word appears less frequently, its inverse document frequency is usually higher, and vice versa.

Therefore, the $T_f-I_{df}$ weights can be expressed as:
 $$
\mathbf{w}_{k j}=t f_{k j} \times i d f_{k}
$$

After the $T_f-I_{df}$ weight is obtained, the cosine similarity of the two product title documents can be calculated. For the specific cosine similarity, see Section 2.2.2. This calculation requires the use of an inverted index. The specific MapReduce process is as follows:

(1) Input the document id, the word segmentation id and the $T_f-I_{df}$ weight of the word segmentation into the map task, use the word segmentation id as the key, and the document id and $T_f-I_{df}$ weight as the value output; (2) In the reduce stage, calculate any pair of documents with the same word segmentation The product of the $T_f-I_{df}$ weights, take the document pair as the key, and the weight product as the value output; (3) Input the original data with the $T_f-I_{df}$ weight to the map task again, take the document id as the key, and the word segmentation $T_f-I_{df}$weight is value output; (4) calculate the modulo of each document in the reduce phase;

(5) For the data output in steps 2 and 4, perform a left outer join with the document as the key; (6) Input the output of the previous step to the map task, and use the document id pair as the key, $T_f-I_{df}$ weight product, The document modulo is the value output; (7) Calculate the cosine similarity of each document pair in the reduce task.
 
 \subsubsection{Ranking correction based on convolutional neural network}
 Convolutional neural network is one of deep learning technologies, and its powerful image processing capability based on GPU has been more and more applied in engineering practice. The specific convolutional neural network is shown in Section 3.4.

This section will use the powerful feature extraction capabilities of convolutional neural networks to extract image features, and then pass them into the logistic regression model for further processing. The process is shown in Figure 3.4.

Firstly, the existing classic neural network model AlexNet is used to extract the image information of each commodity, and the feature is the neuron data of the penultimate layer of the network. From this, the high-dimensional features of the image can be obtained, and then the matching package data, the recall set data and the image feature data are connected respectively, and the difference between the image features of any two products is calculated. Add certain non-package negative sample data to the matching package as the training data for logistic regression. The output model predicts the probability of recalling the set as a matching package, and uses this probability to revise the original ranking.
\begin{figure}[H]
  \centering
  \includegraphics[width=14cm]{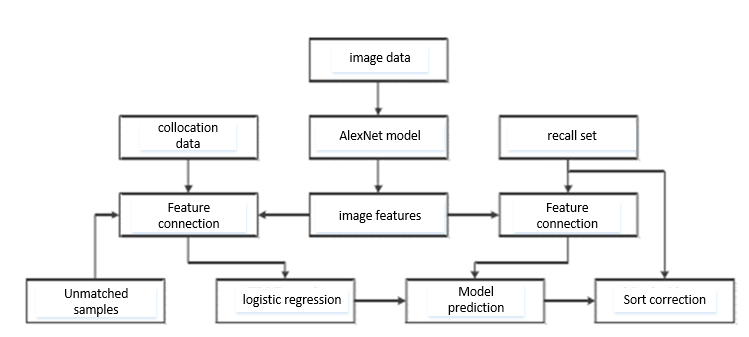}
  \caption{CNN-based ranking correction process}
\end{figure}
 
The AlexNet model was proposed by Alex on the imagenet image classification challenge, and its network parameters are as high as 600 million. The first five layers are all convolutional layers, and the sixth to eighth layers are fully connected layers.

For sorting correction, firstly, the correction degree of each commodity pair needs to be calculated, which can be expressed as:
$$
{ fix }_{i t 1, i t 2}=\frac{p_{i t 1, i t 2}}{\sum\limits_{{pair } \in \operatorname{Re} c a l l} p_{ {pair }}} \times m
$$ 
 
Among them, $p_{i t 1, i t 2}$  is the output frequency of commodity pairs in the recall set in logistic regression, and m is the number of commodity pairs in the whole recall set. After that, the correction degree is standardized, specifically: 
 $$
\operatorname{rank}_{f i x}=f i x_{i t 1, i t 2}-a v g\left(f i x_{p a i r}\right)
$$

Among them,  $avg(\cdot)$  is the mean value. rankfix is the final ranking correction number. In this way, the matching similarity of commodity pairs can be adjusted to a certain extent with the help of image information to a certain extent.
 
\subsection{Experimental results and analysis} 
 \subsubsection{Experimental platform and data}
 All experiments in this section are based on the ODPS platform for big data processing. The total data flow of the experiment exceeds 3PB, including the MapReduce algorithm framework, Caffe deep learning components, common machine learning and natural language processing algorithm packages, and ODPSSql data processing platform.

First, the hybrid-based recommendation system will use the mobile Tmall real data (AliMobileRec) provided by Ali Group. Data provides user ID, product ID, product category , product location information (with missing values), when and where the user performed on the product (click, favorite, add to cart, purchase) and where and when (with a large number of missing values in geographic location information). There are a total of 5822532780 pieces of behavioral data, 524484376 pieces of behavioral data in the vertical field in the complete set of data, and 14397493 pieces of commodity information data. The specific information is shown in Table 3.1 and Table 3.2. The goal is to build a machine learning model to predict which users purchased which items on the forecast day, and then use the model to recommend items.
\begin{table}[H]
\renewcommand\arraystretch{1.5}  
\centering
\caption{Mobile terminal behavior data of the complete set of commodities}
\begin{tabular}{p{3cm} p{3cm} p{8cm} }
\hline
Field Name     & Data Type & Field Meaning                      \\
\hline
User\_id       & bigint    & User ID                            \\
Item\_id       & bigint    & product identification             \\
Behavior\_type & bigint    & Browse, Favorite, Add to Cart, Buy \\
User\_geohash  & string    & latitude and longitude             \\
Item\_category & bigint    & product category                   \\
Time           & string    & behavior time                   \\
\hline
\end{tabular}
\end{table}
 
\begin{table}[H]
\renewcommand\arraystretch{1.5}  
\centering
\caption{Commodity subset data}
\begin{tabular}{p{3cm} p{3cm} p{8cm} }
\hline
 Field Name     & Data Type & Field Meaning                      \\
 \hline
Item\_id       & bigint & product   identification                \\
User\_geohash  & string & Product location space   identification \\
Item\_category & bigint & product category      \\
\hline
\end{tabular}
\end{table}
  
 \begin{table}[H]
\renewcommand\arraystretch{1.5}  
\centering
\caption{Matching packages}
\begin{tabular}{p{3cm} p{3cm} p{8cm} }
\hline
Field Name     & Data Type & Field Meaning                         \\
\hline
Coll\_id      & bigint & with package id   \\
User\_geohash & string & with a list of id  \\
\hline
\end{tabular}
\end{table}
 
  \begin{table}[H]
\renewcommand\arraystretch{1.5}  
\centering
\caption{Commodity Information Sheet}
\begin{tabular}{p{3cm} p{3cm} p{8cm} }
\hline
Field Name     & Data Type & Field Meaning                         \\
\hline
Item\_id  & bigint & product   identification   \\
Cat\_id   & bigint & product category id        \\
Terms     & string & Product title segmentation \\
Img\_data & jpg    & product picture \\
\hline
\end{tabular}
\end{table}
 
   \begin{table}[H]
\renewcommand\arraystretch{1.5}  
\centering
\caption{ User history behavior table}
\begin{tabular}{p{3cm} p{3cm} p{8cm} }
\hline
Field Name     & Data Type & Field Meaning                         \\
\hline
User\_id   & bigint & User   identification  \\
Item\_id   & bigint & product identification \\
Create\_at & string & Purchase date       \\
\hline
\end{tabular}
\end{table}

Secondly, the data set used by the deep learning-based collocation recommendation system is the TaobaoClothesMatch data set. The problem is defined as the set data of some collocation masters, the product information and the user's historical purchase data, and the collocation list of the predicted target id. The purpose is to recommend the matching information of the purchased products to the user based on this kind of clothing matching information, and then associate the sales. The specific data formats are shown in Table 3.3 to Table 3.5.

Among them, bigint, string, and jpg are all stored data structures, which are long-format integers, strings, and compressed image formats, respectively. The scale of the data is 158318 package samples, 3269988 commodity and image information samples, and 728318569 user historical data samples.

\subsubsection{Experimental results}

(1) Experimental results of recommendation system based on hybrid technology

The recommender system will mainly use the F1 indicator for effect evaluation. The F1 index is used here instead of the commonly used accuracy index because the accuracy index generally needs to give a complete test set, and then obtain the correct proportion of the predicted samples. In the current scenario, it is necessary to construct a test set, that is, a recall set. Using the accuracy rate can only express the proportion of correct samples in the constructed recall set, while selecting F1 can express both the proportion of correct samples in the recall set and the number of correct samples in the real samples. The superiority of the system can be reflected only when both are relatively high, so the F1 index is selected as the evaluation index of the current system.

The data set gives the data from November 18th to December 18th, because the full set of verification data on the 19th cannot be obtained, so the data on December 18th is used as the verification set, and the number of submissions is 105,000 user-goods pairs. A total of 1200 multi-dimensional features are constructed on the ODPS platform. See Section 3.2.2 for the feature construction method.

First, verify the relationship between the number of features and the system effect, using the GDBT model, the positive and negative sampling ratio is 1:8, the model parameters are metric type AUC, the number of trees is 500, the learning rate is 0.05, the maximum number of cotyledons is 32, the maximum tree depth is 11, and the leaves The minimum number of samples for nodes is 500, the proportion of samples collected for training is 0.6, the proportion of features collected for training is 0.6, and the maximum number of splits is 500. The effect is shown in Figure 3.5.
\begin{figure}[H]
  \centering
  \includegraphics[width=12cm]{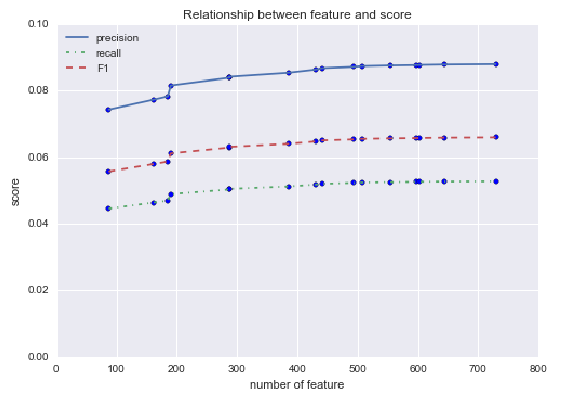}
  \caption{The relationship between the number of features and the score}
\end{figure}

Among them, the added feature categories are: time of day and shopping cart features, user features, user product subset features, product subset features, product complete set features, product category features, product competition features, user product category features, and user product complete set features, etc.

Verify the effect of feature-level fusion on the recommendation effect. Due to the limitation of the algorithm platform, only a maximum of 799-dimensional features can be used. Therefore, three sets of features are randomly selected, each with 799 dimensions, and the GBDT model is used for training, and the parameters are the same as above. The specific effects are shown in Table 3.6.
\begin{table}[H]
\renewcommand\arraystretch{1.5}  
\centering
\caption{Feature-level fusion effect table}
\begin{tabular}{p{6cm} p{3cm} p{3cm}  p{3cm}}
\hline
feature set                               & Precision   & Recall      & F1          \\
\hline
feature set 1                             & 0.090503845 & 0.054395265 & 0.067950461 \\
feature set 2                             & 0.088748303 & 0.053772593 & 0.066968796 \\
feature set 3                             & 0.091227602 & 0.054738021 & 0.068421842 \\
feature set 1+feature set 2               & 0.089557052 & 0.053938258 & 0.067326959 \\
feature set 2+feature set 3               & 0.089729976 & 0.054063935 & 0.067473728 \\
feature set 1+feature set 3               & 0.091543545 & 0.055075063 & 0.068773897 \\
feature set 1+feature set 2+feature set 3 & 0.091087625 & 0.05484656  & 0.068467068\\
\hline
\end{tabular}
\end{table}
\begin{table}[H]
\renewcommand\arraystretch{1.5}  
\centering
\caption{Effect table of different filtering and feature selection methods}
\begin{tabular}{p{6.5cm} p{2.5cm} p{3cm}  p{3cm}}
\hline
Feature or sample filtering method       & Precision   & Recall      & F1          \\
\hline
Linear correlation filter feature        & 0.091638095 & 0.054966524 & 0.06871581  \\
LR filtered samples                      & 0.09212381  & 0.055257866 & 0.069080028 \\
Crawler rules filter samples             & 0.09067619  & 0.054389553 & 0.067994515 \\
Random Negative Sampling Filters Samples & 0.090685714 & 0.054395265 & 0.068001657 \\
Linear correlation + LR filtering        & 0.092590476 & 0.055537783 & 0.069429963\\
\hline
\end{tabular}
\end{table}
\begin{table}[H]
\renewcommand\arraystretch{1.5}  
\centering
\caption{ Effect table of different recall sets}
\begin{tabular}{p{6cm} p{3cm} p{3cm}  p{3cm}}
\hline
Recall Set                        & Precision   & Recall      & F1          \\
\hline
Cart recall set                   & 0.08552381  & 0.051299043 & 0.064130947 \\
shopping cart + non-shopping cart & 0.090742857 & 0.054429541 & 0.068044506\\
\hline
\end{tabular}
\end{table}
\begin{table}[H]
\renewcommand\arraystretch{1.5}  
\centering
\caption{Fusion effect table of different random negative sampling}
\begin{tabular}{p{6cm} p{3cm} p{3cm}  p{3cm}}
\hline
negative sampling set      & Precision   & Recall      & F1          \\
\hline
Random Negative Sampling 1 & 0.090780952 & 0.054452391 & 0.068073072 \\
Random Negative Sampling 2 & 0.090666667 & 0.05438384  & 0.067987374 \\
Random Negative Sampling 3 & 0.090619048 & 0.054355277 & 0.067951666 \\
Three sets of fusion       & 0.090952381 & 0.054555218 & 0.06820162 \\
\hline
\end{tabular}
\end{table}

To verify the influence of different filtering methods and feature selection methods on the system, the GDBT model is used, the parameters are the same as above, and the effect is shown in Table 3.7. To verify the impact of different recall sets and weighted fusion on the system, the GDBT model is used, the parameters are the same as above, and the effect is shown in Table 3.8. Among them, weighted fusion refers to the use of a hierarchical mixing strategy, first taking certain data from the recall set with high hits, and then taking certain data from another recall set when the hit is low to a threshold, thereby ensuring the final selection of multiple recall sets. The data hit rate is comparable. In this experiment, after verification and tuning, 88,500 samples were taken from the shopping cart recall set and 16,500 samples were taken from the non-shopping cart recall set during fusion.

To verify the impact of different model fusions on the system, the GDBT model is used, and the parameters are the same as above. The difference between the models is mainly due to different random negative sampling, and the effects are shown in Table 3.9.

In the end, the online (validated on December 19) F1 value obtained by the fusion of different models was 8.11\%.

The above experimental results show that:

$\textcircled{1}$ With the increase of features, the model effect will gradually improve. The effect of the improvement is large at the beginning, and then the effect is gentle, but the generalization error does not decrease. This is why the sliding window sampling is used to increase the number of samples, making the model less likely to overfit , and the features added afterward have not been greatly improved, indicating that a lot of feature information has been redundant.

$\textcircled{2}$ Fusion when the features are good enough is helpful to improve the system effect, but when some single models perform poorly, the fusion will reduce the effect of the system.

$\textcircled{3}$ When the feature has many redundant features, the linear correlation feature selection will greatly improve the system effect because of the redundant features. Using the waterfall hybrid strategy, that is, using the model for data filtering, performs well.

$\textcircled{4}$For recall sets with different hit rates, it is not possible to simply fuse them on average. Instead, a hierarchical hybrid strategy should be used, and different fusion weights should be set according to the hit rate to achieve the effect of improving the model.

$\textcircled{5}$ When the effect of the single model reaches the standard, the fusion model will get a certain improvement in effect, but if the difference between the models is not large, the effect will not be significantly improved.

(2) Experiment results of collocation recommendation system based on deep learning

This experiment uses the MAP index as the evaluation index. This indicator is selected because the collocation set is a relatively unpredictable set, and if the accuracy rate is selected as the evaluation standard, the same problem as the previous recommendation system will occur. If the F1 indicator is selected, the score is likely to be zero or very small, which cannot be Reflect the real system effect. Therefore, each product needs to provide 200 alternative products as its matching products. Here, it is necessary to take care of the accuracy of the recommendation, and to make the accurate products as far as possible to predict in front. Due to the existence of these problems and needs, choose MAP here. indicators as evaluation indicators.

The feature extraction model uses the AlexNet model. Reorder the probability output using a logistic regression model. The logistic regression model parameters are the maximum number of iterations 1000, the convergence error is 0.000001, the L1 is regular, and the coefficient is 1.

First, use the matching package set to evaluate the classification performance of the features extracted by the convolutional neural network and the logistic regression model. The evaluation method uses the cross-validation technique, and a part of the samples are reserved as the validation set. The validation set does not participate in the model training, and the results are shown in the table. 3.10. The experimental results of each part of the whole recommendation system are shown in Table 3.11. The final online MAP value of this experiment was 4.66\%.

The experimental results show that when the samples of the verification set are balanced, the features extracted by the convolutional neural network will produce a better classification effect in the logistic regression model, especially when the training results are close to the verification results, which indicates that the model is not overfitting. Generally speaking, these features have a strong ability to describe products, but when used in the whole system, although the improvement is not large, the main reason for this phenomenon is that the hit rate of the recall set is too low, and the hit rate is too low. Higher recall set triggering techniques should improve system performance.
\begin{table}[H]
\renewcommand\arraystretch{1.5}  
\centering
\caption{CNN network extraction feature effect table}
\begin{tabular}{p{3.5cm} p{3.5cm} p{3.5cm}  p{3.5cm}}
\hline
                  & Correct rate & Recall rate & F1          \\
                  \hline
Validation results & 0.649624425  & 0.250524847 & 0.361600158 \\
training results   & 0.650327156  & 0.256068934 & 0.367452118\\
\hline
\end{tabular}
\end{table}
\begin{table}[H]
\renewcommand\arraystretch{1.5}  
\centering
\caption{The effect table of each part of the recommendation system}
\begin{tabular}{p{7.5cm} p{7.5cm}}
\hline
 method                                & MAP        \\
 \hline
User history recall set               & 0.02586574 \\
Commodity tfidf similarity recall set & 0.02506066 \\
Fusion of two recall sets             & 0.04474209\\
CNN correction	&0.0466117\\
\hline
\end{tabular}
\end{table}

\section*{References}
[1] Ma Yani. Design and implementation of scoring system based on intelligent recommendation model [J]. Microcomputer Application, 2019, 35(03): 70-72.

[2] Yu Wei, Xu Dehua. Overview and Prospect of Recommendation Algorithms [J]. Science and Technology and Innovation, 2019, (04): 50-52.

[3] Lin Jianhong. Analysis of the application of artificial intelligence technology in the field of e-commerce [J]. China Business Review, 2019, (02): 19-20.

[4] Huang Maozhou. Algorithms in Recommendation Systems [J]. Technology Wind, 2019, (01): 22-23.

[5] Wang Junshu, Zhang Guoming, Hu Bin. A Review of Recommendation Algorithms Based on Deep Learning [J]. Journal of Nanjing Normal University (Engineering Technology Edition), 2018, 18(04): 33-43.

[6] Zhang Teng, Lin Guimin, Qiu Lida, Liu Chaoming, Wei Yujing. Traffic data analysis and application based on machine learning [J]. Modern Information Technology, 2018, 2(12): 16-18.

[7] Chen Guo, Zhou Zhifeng, Yang Xiaobo, Wang Cheng, Ouyang Chunping. Design and implementation of a product recommendation system based on face recognition [J]. Computer Age, 2018, (11): 52-55.

[8] Shi Hongyi. Overview of Machine Learning [J]. Communication World, 2018, (10): 253-254.

[9] Lin Haining. Recommendation system using Google Tensorflow [J]. Computer Knowledge and Technology, 2018, 14(28): 195-196.

[10] Zhong Wei, Li Zhichen. Research on Network Education System Based on Machine Learning [J]. Journal of Communications, 2018, 39(S1): 135-140.

[11] Li Jinzhong, Liu Guanjun, Yan Chungang, Jiang Changjun. Research Progress and Prospects of Ranking Learning [J]. Chinese Journal of Automation, 2018, 44(08): 1345-1369.

[12] Zhang Quangui, Li Zhiqiang, Zhang Xinxin, Cao Zhiqiang. Recommendation of Deep Joint Learning Integrating Metadata and Attention Mechanism [J]. Computer Application Research, 2019, (11): 1-3.

[13] Kuang Wenbo, Tong Wenjie. On the algorithm model design of video intelligent recommendation [J]. News and Communication, 2018, (15): 4-9.

[14] Xiao Shibo, Guo Xiuying. Research on document personalized recommendation system based on user characteristics [J]. Network New Media Technology, 2018, 7(04): 24-33.

[15] Lin Fei, Zhang Zhan. A heap model based on downsampling to predict the learning outcomes of large-scale network courses [J]. Computer Applications and Software, 2018, 35(07): 131-137.

[16] Wang Xidian, Wang Lei, Long Quan, Xue Yang. Artificial intelligence and its application in network optimization operation and maintenance [J]. Telecommunications Engineering Technology and Standardization, 2018, 31(07): 81-86.

[17] Lu Langlang, Yuan Qingda, Ling Yuanjun, Lin Yunpeng, Wang Haoyu. Design of Intelligent Recommendation Algorithm Based on Latent Semantic Model [J]. Science and Technology Communication, 2018, 10(13): 124-126

\end{spacing}   


\end{document}